# Behavior of vortices near twin boundaries in underdoped Ba(Fe$_{1-x}$Co$_x$)$_2$As$_2$

B. Kalisky[1,2,*], J.R. Kirtley[1,2], J.G. Analytis[1,2,3], J.-H. Chu[1,2,3], I.R. Fisher[1,2,3], K.A. Moler[1,2,3,4,*]

[1] *Geballe Laboratory for Advanced Materials, Stanford University, Stanford, California 94305-4045, USA*
[2] *Department of Applied Physics, Stanford University, Stanford, California 94305-4045, USA*
[3] *Stanford Institute for Materials and Energy Sciences, SLAC National Accelerator Laboratory, 2575 Sand Hill Road, Menlo Park, CA 94025*
[4] *Department of Physics, Stanford University, Stanford, California 94305-4045, USA*
* Correspondence to: Beena Kalisky (beena@stanford.edu); Kathryn A. Moler (kmoler@stanford.edu)

We use scanning SQUID microscopy to investigate the behavior of vortices in the presence of twin boundaries in the pnictide superconductor Ba(Fe$_{1-x}$Co$_x$)$_2$As$_2$. We show that the vortices avoid pinning on twin boundaries. Individual vortices move in a preferential way when manipulated with the SQUID: they tend to not cross a twin boundary, but rather to move parallel to it. This behavior can be explained by the observation of enhanced superfluid density on twin boundaries in Ba(Fe$_{1-x}$Co$_x$)$_2$As$_2$. The observed repulsion from twin boundaries may be a mechanism for enhanced critical currents observed in twinned samples in pnictides and other superconductors.

Dissipation from moving vortices is a major limitation for the use of superconductors in high current density applications. Vortex motion in superconductors is controlled by competition between the Lorentz force in the presence of an applied current, thermal energy, the interactions between vortices, and their interaction with the local pinning landscape [1-5]. Sources of pinning locally reduce the free energy of a vortex that passes though them, resulting in improved critical current in samples with strong pinning [1, 6-10]. The various types of pinning sources, including oxygen vacancies, impurities, dislocations, extended columnar defects caused by irradiation, and others [1, 11-15], can be broadly categorized by the dimensionality of the pinning potentiality. One technologically important two-dimensional pinning source is grain boundaries, the boundaries between grains with different crystalline orientations [16]. If the angle between the two grains is low, dislocations are formed, separated by regions that are well lattice-matched. In this case, the lattice-matched area remains superconducting and the dislocations act as pinning sites. At high angles the misorientation is hard to accommodate, the superconductivity is reduced, and the grain boundary behaves like a Josephson junction. Twin boundaries are a particular kind of grain boundary formed in materials with orthorhombic symmetry by a nearly 90 degree rotation around the *c*-axis, or equivalently by an interchange between the *a* and *b* crystalline axes. Twin boundaries can be found in many of the relatively new family of pnictide superconductors [17, 18]. The 122 pnictide compounds undergo a tetragonal to orthorhombic transition and twin formation occurs prior to entering the superconducting phase in the underdoped part of the superconducting dome [19-21]. We have previously reported enhanced superfluid density on twin boundaries in the Cobalt-doped 122 pnictide family [22, 23], while bulk magnetization measurements show enhanced vortex pinning associated with the presence of twins [24].

In the present work we wish to examine the effect of twin boundaries in pnictides on the vortices and their dynamics. We use scanning Superconducting QUantum Interference Device (SQUID) microscopy, which is local and sensitive enough to measure both individual vortices and the local variations in the superfluid density on twin boundaries [22, 23]. We report that *vortices avoid twin boundaries*. We also report that twin boundaries act as strong barriers for vortex motion, which may cause enhancement of the critical current.

Twin boundaries are common in orthorhombic crystals of other families of superconductors, such as Sn (type 1) and Nb (type 2), and earlier work from bulk measurements indicates that the superconducting properties are different at the twin boundaries [25]. Studies of the effect of twin boundaries on vortex motion in conventional superconductors indicate pinning on twin planes [26]. Twin boundaries in the cuprate high-$T_c$ superconductor $YBa_2C_3O_x$ (YBCO) have been extensively studied and the twin boundaries are considered strong pinning sites, as evidenced by the increase in critical current in twinned samples [7-10, 27], magnetization measurements [28] and direct observation of vortices lining up on the twin boundaries [29-33]. Magneto-optical studies of flux penetration show that, depending on the direction the flux flows, a twin boundary can behave as a channel for easy vortex motion [34] or as a barrier to vortex propagation [35]. Both behaviors are related to the buildup of flux at the boundary due to its stronger pinning [36], which is attributed to *reduced* superfluid density. Our contrasting observation of *enhanced* superfluid density on twin boundaries in the Cobalt doped 122 pnictide family motivates the present study.

The growth of the crystals used in this study is described in Ref. [20]. In the following we show data from two underdoped (x=5.1%) $Ba(Fe_{1-x}Co_x)_2As_2$ samples (S1 and S2). The critical temperatures measured for these samples are 18.25±0.25K for the superconducting transition $T_C$ (measured in situ), and 36.8±7K and 55±5K for the magnetic and structural transitions (determined by the temperature derivatives of resistivity as in Ref. [20]). We used a variable-*T* scanning SQUID susceptometer [37-39] with two field coil / pickup loop pairs in a gradiometer configuration. Figure 1a shows the sensing region of the probe, which was polished to a corner to bring it as close as possible to the sample. Two types of measurements were performed with the SQUID: (1) Magnetometry, in which the flux through the pickup loop is recorded while the SQUID is scanned over the surface of the sample. The flux measured is the local magnetic field of the sample integrated over the pickup loop area (shown in units of $\Phi_0$); (2) Susceptometry, in which an ac current in the field coil generates a local magnetic field, and the resulting total ac flux through the SQUID pickup loop is recorded while scanning. The flux measured includes the magnetic response of the sample to the locally applied field, and is normalized by the current in the field coil (shown in units of $\Phi_0$/A). The samples' response to the locally applied field, which is measured by SQUID susceptometry, can be directly related to the local magnetic penetration depth and superfluid density [23].

In our previous work we observed that the susceptometry signal is spatially modulated in underdoped samples of $Ba(Fe_{1-x}Co_x)_2As_2$ that also show twinning, and concluded that the superfluid density is enhanced on twin boundaries in this material. Figure 1b shows a susceptometry scan of the surface of an underdoped (x=5.1%) $Ba(Fe_{1-x}Co_x)_2As_2$ sample (S1). The whiter diagonal lines, where the susceptibility signal is enhanced, represent the locations of the twin boundaries. The mottled background is associated with the local variations of the critical temperature, which are observable in susceptibility images taken near $T_C$.

Measuring the same area by SQUID susceptometry and SQUID magnetometry provides a unique opportunity to follow how the vortex behavior is modified by the twin boundaries. Such data can be taken either simultaneously or by successive images of each scan type taken over the same region under identical conditions. An overlay of susceptometry (Fig. 1b) and magnetometry (Fig. 1c) scans of the same region in sample S1 is shown in Figure 1d. The magnetometry image shows individual pinned vortices. The applied magnetic field in which the sample was cooled has been adjusted to produce a low density of vortices upon cooling through $T_C$. These images are similar to those observed in other type II superconductors, except that on close inspection it appears that every vortex avoids pinning directly on a twin boundary in multiple thermal cycles on multiple samples. The susceptometry data is shown close to $T_C$, where the twin boundary susceptibility signal is the largest. The magnetometry data is shown at 4.8K since the vortices are smaller and their centers are more easily determined at low temperatures. That the vortices are found only between the twin boundaries is in sharp contrast to the behavior observed in YBCO, where vortices decorate twin boundaries [29-33], and $ErNi_2B_2C$, where, in a small magnetic field, all vortices appear to sit on twin boundaries in the antiferromagnetic and superconducting state [40, 41].

As displayed in Fig. 1d, the vortex distribution appears close to a uniform random distribution, but the vortices appear to avoid twin boundaries. Figure 2 shows the distribution of the distance between each vortex and its nearest twin. This data was acquired by imaging the same region multiple times after thermal cycling above the superconducting transition temperature $T_c$ to pin vortices in new positions, but well below the structural transition temperature (55K±5K) so that the twin boundaries remained at the same location [22]. We also show the normalized histogram expected for a uniform random distribution, generated by calculating the distance to the nearest twin from every pixel in the image for our twin boundary configuration (Fig. 1b). This histogram would be flat for distances less than half the smallest twin boundary spacing for a large area with uniformly spaced twins. The overlay of the two normalized histograms shows that vortices are very unlikely to be found within a micron of a twin boundary.

Although at first glance the vortices appear to be pinned in random locations aside from their avoidance of sitting directly on a twin boundary, on closer inspection we notice that vortices are more likely to be closer to the twin boundary on their right than the twin boundary on their left. It also appears that the vortices are slightly more likely to be found within 1-7 microns of a twin boundary than within 9-13 microns of a twin boundary, even after accounting for the variable spacing between twin boundaries. In order to test whether the two histograms for the experimental pinned positions and random pinning are significantly different, we used the Kolmogorov-Smirnov test [42]. The resulting p-value (i.e. the probability that the two samples were drawn from the same distribution) is 6%. Given the mostly random pinning throughout the bulk of this sample, additional statistics would be needed to make a definitive statement as to whether there is also an attractive interaction between vortices and twin boundaries at length scales of 1-7 microns.

Forty of the sixty vortices that were imaged on the twins configuration in Fig. 1b were found closer to the twin to their right (see *e.g.* Fig. 1d.). Such a trend could be explained by the proximity of the measured area to an edge of the sample. For example, Meissner forces would tend to push the vortices towards the outside of the sample upon cooling, and vortices could be preferentially pinned close to the sides of the twins nearest the sample edge.

The enhancement of superfluid density on the twin boundaries should theoretically result in an increased energy for a vortex that is near enough to a twin boundary that its current profile overlaps with the region of enhanced superfluid density, and therefore should result in repulsive interaction between the vortices and the twin boundaries [23, 43]. Such an effect would decrease the probability of finding vortices near the twin boundary. In the presence of a strong pinning landscape throughout the bulk of the sample, a vortex repelled from the twin boundary would likely find itself pinned when the vortex-twin boundary interaction is not strong enough to overcome the effect of the strong pinning landscape, explaining the peak in the second bin of the histogram in figure 2.

The ability to manipulate vortices is very useful for the study of pinning effects in superconductors [44, 45], and the tendency of vortices to avoid twin boundaries becomes even more obvious when we examine their dynamics under the influence of a locally applied driving force. Figure 3 shows magnetometry (Fig. 3a-b) and susceptometry (Fig. 3c-d) images taken over the same region of another x=5.1% sample (S2) at two temperatures. At 5K (3a,c) the vortices do not move during the imaging process. Comparing their positions to the twin boundary locations [46], where these boundaries are spaced far enough apart to be resolved, confirms that the vortices do not pin on twins, as discussed above. The vortices in Fig. 3a seem to be correlated in a way that might suggest a tendency towards vortex pairing, although such correlations are difficult to distinguish from a clumping tendency that might result from a mesoscopically disordered strong local pinning landscape [47]. At the higher temperature of 15K (Fig. 3b,d) the vortices are dragged by the small local field applied by the SQUID sensor. The fast scan direction in these images is horizontal (as indicated by the double arrows in Fig. 3b), and on average the vortices are expected to be dragged vertically, as observed in other materials. However, most of the vortices move diagonally, parallel to the twins. One example of a vortex moving parallel to a twin is circled in Fig. 3b. The dashed line plotted along its motion path in Fig. 3b is replotted at the same position on Fig. 3d. This vortex is close to the twin observed to its right and moves parallel to it.

The field coil on our SQUID sensor can be used as a vortex manipulation tool [44] because a current though our field coil exerts quantitatively calculable, measurable, and controllable forces on a vortex during imaging. The SQUID sensor can exert forces on vortices in several ways. First, and primarily, the field coil is designed to carry a current, the field from which exerts a force on any nearby vortices [44]. Second, the d.c. component of the current through the pickup loop, which is comparable to the critical current of the SQUID, exerts a small force on vortices while scanning. Third, there are a.c.

currents through the pickup loop due to the a.c. Josephson effect in the SQUID. The role of these a.c. currents in dragging vortices may be small, because their magnitudes are estimated to be approximately 0.1µA and because the forces they exert tend to average to zero. The locally applied fields from the field coil and the pickup loop induce screening currents in the sample, thereby exerting Lorentz forces on the vortex.

To calculate the force on a vortex, we first need to calculate the magnetic fields within the sample due to the field coil and pickup loop. This can be done based on the method of Kogan *et al.* [48], which involves solving the London equations at the surface of a semi-infinite superconductor by Fourier transforms. The screening currents are determined by Maxwell's equation $\vec{J}=\vec{\nabla}\times\vec{H}$, and the Lorentz force on the vortex core is $\vec{F}=\Phi_0\int_{-\infty}^{0}\vec{J}\times\hat{z}\,\mathrm{d}z$. The force is integrated along the length of the vortex, which is assumed to be a rigid cylinder. The force in the radial direction on a vortex a distance *r* from the axis of the current loop is given by

$$(1) \quad F_r(r)=-\frac{\Phi_0\, I_{loop}\, r_{loop}}{\lambda_{ab}^2}\int_0^\infty dk\,\frac{ke^{-kh}}{q(q+k)}J_1(kr_{loop})J_1(kr)$$

where $r_{loop}$ and $I_{loop}$ are the radius and current through the loop, $q=\sqrt{\lambda_{ab}^{-2}+k^2}$, $\lambda_{ab}$ is the in-plane penetration depth, $\Phi_0$ is the superconducting flux quantum, and *h* is the height of the current loop above the sample. In Figure 4a we plot the in-plane force applied by the field coil (black dashed) and the pickup loop (red solid) for typical scan values for S1-S2 at 5K: *h*=1µm, $\lambda_{ab}$=340nm, $r_{loop}$=10µm and $I_{loop}$=250µA for the field coil, and $r_{loop}$=1.5µm and $I_{loop}$=10µA for the pickup loop. In Figure 4b we plot the calculated force curves for various experimental situations. From the black dashed curve in Fig. 4b, we infer that the vortex in Fig 3e experienced an applied force of up to 2pN applied by the field coil. Despite the applied force, this vortex did not cross the twin boundary but moved easily along it.

Sufficiently close to $T_c$, the vortices are large and very mobile, such that scanning the SQUID drags the vortices easily. In this case in some regions the vortices are observed decorating the twin boundaries locations. For example, Fig. 5 shows a section of the same region imaged in Fig. 1 at 18K. Stripes in the magnetic flux through the SQUID are observed although there is no current through the field coil, and we attribute this flux to vortices in motion.

The SQUID provides an independent view of the susceptibility contrast that identifies the twin boundary locations. We use magnetometry to locate vortices within a reasonable dragging distance from the known twin location, and further investigate the strength of the barrier that was set by the twin boundary. Figure 6 shows vortices being dragged across a twin boundary. In order to gradually increase the effectiveness of the dragging force applied by the SQUID, magnetometry scans were taken at increasing temperatures. The magnetometry data, showing two vortices, is superimposed on susceptometry data of the same area to illustrate the location of the twin boundaries. The vortex appears to cross the twin in the middle of the 17K scan (marked by an arrow on Fig. 6c). As shown in Fig. 4, the force applied by the pickup loop is much weaker than the force applied by the field coil, which typically carries current which is 100 times larger [44]. For the pickup loop parameters: current (10µA), radius (1.5µm), scan height above the sample (1µm), and penetration depth ($\lambda_{ab}$(17K)~740nm [47]), we estimate the force applied by the pickup loop at this scan as $1.6\times10^{-3}$ pN at 17K (see the blue dotted curve in Fig. 4b). For the right hand vortex, the dragging was possibly done by the leads of the pickup loop which, depending on the specific alignment, may be closer to the sample's surface (see arrow in Fig. 6c) [49].

Few of the vortices were found very close to the twin boundary (see Fig. 1d). Due to our limited spatial resolution for one of the vortices it was hard to determine whether it is trapped on the boundary or very close to it. Dragging experiments can help determine that. A vortex that is dragged in various directions is expected to reveal a barrier for both dragging "into" and "away" from the twin boundary if the vortex is trapped *on* the twin boundary. However, in our case, dragging the vortex in different directions shows a barrier for the vortex motion only when dragging "into" the presumed twin, confirming that the vortex is pinned outside of the twin boundary.

The observation that vortices avoid pinning on the twin boundaries is consistent with the observation of enhanced superfluid density on the twins: it is more energetically costly to create a vortex in a region of enhanced superfluid density. A repulsive force is expected from the enhanced superfluid density on the twin boundary. This force is maximal when the vortex approaches the twin boundary and depends on the extent of the superfluid density variation across the twin boundary [43].

In YBCO, Bitter decoration experiments observed vortices lining up on the twin boundaries [29-33], which led to classifying twin boundaries as locations with stronger pinning. However, some of the magneto-optics works that followed the penetration of vortices to the sample near twin boundaries reported that the twins act as barriers for vortex motion, and suggested that the vortices accumulate *near* the twin boundary and not on it [35]. It would be interesting to measure the local superfluid density and vortex behavior near twin boundaries in YBCO and determine whether the twin boundaries in YBCO have lower superfluid density, leading to accumulation of vortices on the twins, or higher superfluid density like in the pnictides.

The possibility of higher superfluid density on twin boundaries in YBCO was discussed by Abrikosov and Buzdin in Ref. [43] and reconciled with the vortex behavior by introducing an attractive force between the vortices and twin boundaries in addition to the repulsion due to the enhanced superfluid density. The source of this attraction is the reduced transparency of many types of grain boundaries in cuprate superconductors [16, 50]. If the twin plane has reduced transparency, the vortices will be attracted to the twin plane similarly to their attraction to the sample edge, in the Bean-Livingston surface barrier description [51]. Such attraction force might be balanced by the repulsive force only very close to the twin, explaining the picture in YBCO. The transparency of twin boundaries in the pnictides is still unknown. Further investigation of the vortex distribution near twin boundaries could gain information about that, although pinning must also be considered.

The barrier for vortex motion realized by the twin boundaries provides a mechanism for enhanced critical currents in twinned samples. In the pnictides such enhancement was observed by Prozorov *et al.* [24] and associated with pinning on the twin boundaries. We suggest that the twin boundaries, which serve as barriers for vortex motion, can explain the same data. In YBCO the observation of stronger critical currents [7-10, 27] was attributed to pinning on the twin boundaries. As discussed above, we suggest that in the pnictides the enhancement of the critical current is due to the barrier placed by the twin boundaries and not due to a potential well that traps vortices on the twin boundaries. Since twin boundaries typically form in both parallel and normal orientations, we expect vortices to be caged between them. This can contribute to enhancement of the critical current as discussed in Ref. [52]. Unfortunately it is not possible to estimate the critical current enhancement from our existing datasets since it requires good statistics of dragging many vortices across many twin boundaries.

In summary, in this work we used scanning SQUID microscopy to investigate the behavior of vortices in the presence of twin boundaries in $Ba(Fe_{1-x}Co_x)_2As_2$. We show that vortices avoid pinning on twin boundaries, but are more likely to be found in proximity to them. Dragging individual vortices shows that the vortices tend to move parallel to the twin boundaries and not to cross them. Our main observation, that twin boundaries are a barrier for vortex motion, provides a mechanism for the enhancement of critical currents. Our observation, that vortices do not pin on the twin boundaries, seems to be in contrast to the pinning that was observed on the twin boundaries in the cuprate superconductors in Bitter decoration experiments [29-33]. Similar to Ref. [35], we suggest that the vortices are not trapped on the twin boundary. The most likely explanation for the difference in the proximity of the vortices to the twin boundary in the cuprates *vs*. the pnictides could be the observed difference between the superfluid change on the twin boundary (reduction vs. enhancement) [22, 23]. This behavior provides a mechanism for enhancement of the critical current in these materials.

**Acknowledgements:** The authors acknowledge many useful discussions with O.M. Auslaender and L. Luan. This work was supported by the Department of Energy DOE under Contract No. DE-AC02-76SF00515, by NSF under Grant No. PHY-0425897, and by the U.S.-Israel Binational Science Foundation. B.K. acknowledges support from L'Oreal USA for Women in Science. J.R.K. was partially supported with a Humboldt Forschungspreis.

**Figure 1:** Pinning positions of vortices in the presence of twin boundaries. **(a)** Optical micrograph of the sensing area of the scanning SQUID susceptometer, showing a 3μm diameter pickup loop surrounded by a field coil **(b)** Susceptometry and **(c)** magnetometry images of the same region of sample S1. **(d)** Overlay of (b) and (c). This overlay shows that the vortices do not pin on the twin boundaries.

**Figure 2:** Distribution of distances between the vortex positions and the nearest twin boundary in the area of Fig. 1. The blue histogram shows the distribution of the distances from the nearest twin for random pinning. This distribution is not flat because the twins are not uniformly spaced. The red shaded bars show the distribution of the distance from the nearest twin for vortices that were imaged with this particular twin configuration. The count is normalized to the total number of vortices measured in this region (60 vortices).

**Figure 3:** Vortex motion near twin boundaries. Magnetometry images showing vortices **(a-b)** and susceptometry images showing twin boundaries **(c-d)** taken on the same region of sample S2 at 5K (a,c) and 15K (b,d). At 5K the vortices do not move during scanning. At 15K the vortices are dragged by an interaction with the SQUID sensor. They tend to move parallel to the twin boundaries, and to not cross them. A force up to ~2pN was applied on the vortex shown in **(e)** by driving current through the field coil, but still the vortex did not cross the twin but rather traveled along it. The arrows in (b) denote the scan directions: the SQUID position is incremented along the single long arrow (slow scan direction) and moves back and forth parallel to the double arrows (fast scan direction).

**Figure 4:** Calculations of the in-plane dragging force applied on a vortex by the SQUID field coil and the SQUID pickup loop. The number in the legend indicates by how much the signal was multiplied in order to plot it on this scale. **(a)** Plotted for a scan height $h$=1μm, $\lambda_{ab}$ = 340nm: Dashed black- SQUID field coil, $r_{loop}$=10μm, $I_{loop}$=250μA; Solid red - SQUID pickup loop, $r_{loop}$=1.5μm, $I_{loop}$=10μA. **(b)** Calculations of forces with parameters relevant to particular images in this paper: Dashed black (Fig.3e)- SQUID field coil, $r_{loop}$=10μm, $I_{loop}$=5mA, $\lambda_{ab}$ = 450nm, $h$=1μm; Solid red (Fig. 6c)– SQUID pickup loop, $r_{loop}$=1.5μm, $I_{loop}$=10μA, $\lambda_{ab}$ = 740nm, $h$=1μm.

**Figure 5:** SQUID magnetometry at 18K (close to $T_C$=18.25K±0.25K) of an area containing twin boundaries in sample S1. The area shown is a portion of the area in Fig. 1a.

**Figure 6:** Dragging a vortex across a twin boundary. SQUID Magnetometry images (jet) show two vortices imaged at increasing temperatures: 4.8K **(a)**, 12K **(b)**, 17K **(c)**, 17.5K **(d)**. These images are superimposed on a susceptometry scan of the same location in S1 taken at 17K (gray scale).

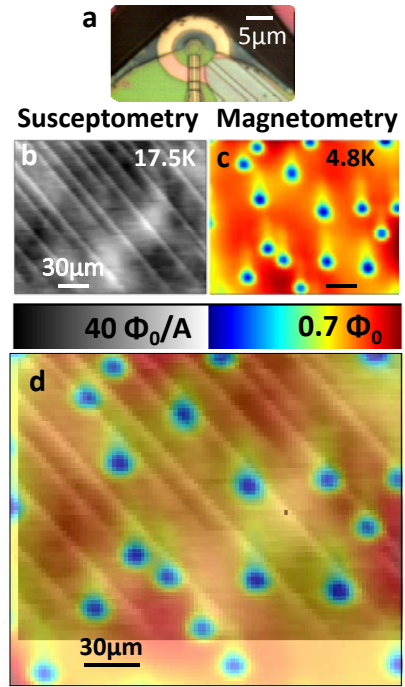

Figure 1

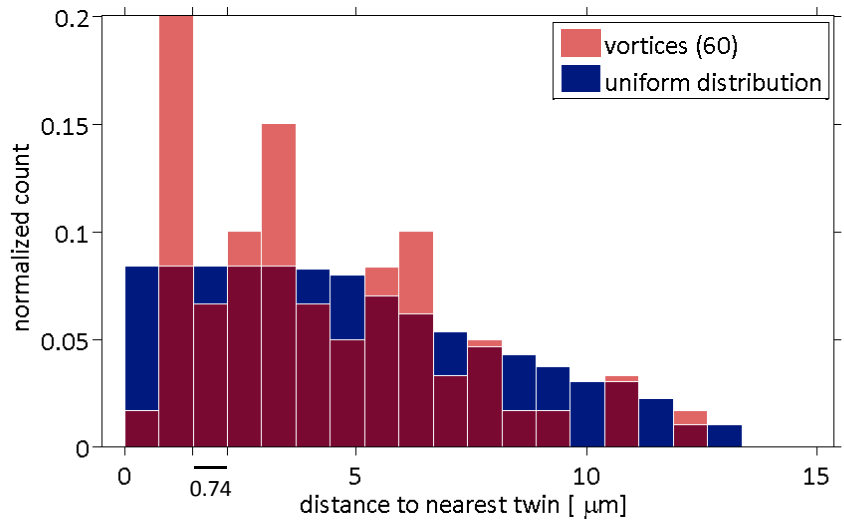

Figure 2

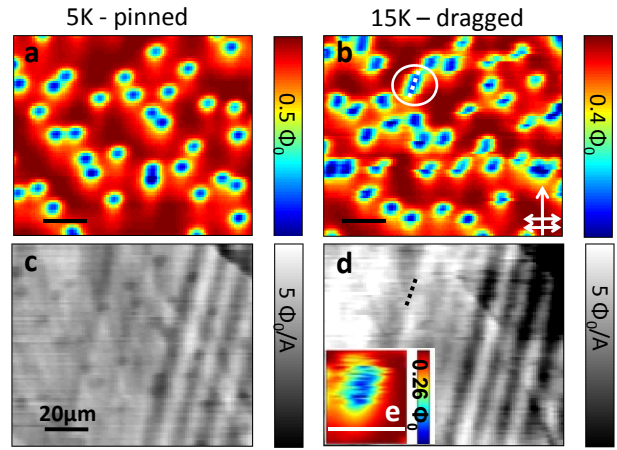

Figure 3

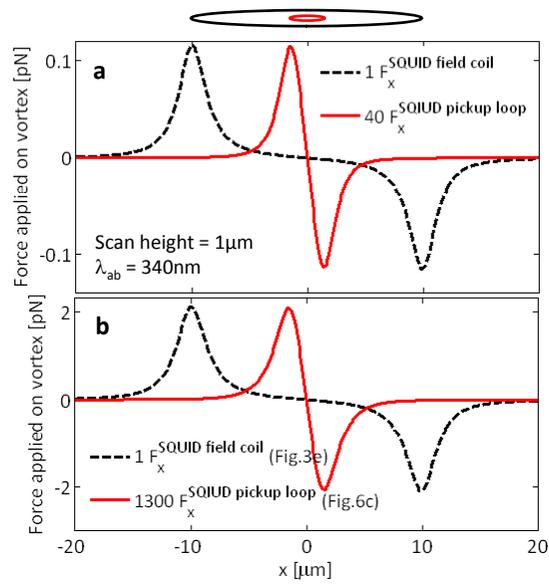

Figure 4

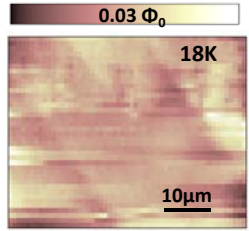

Figure 5

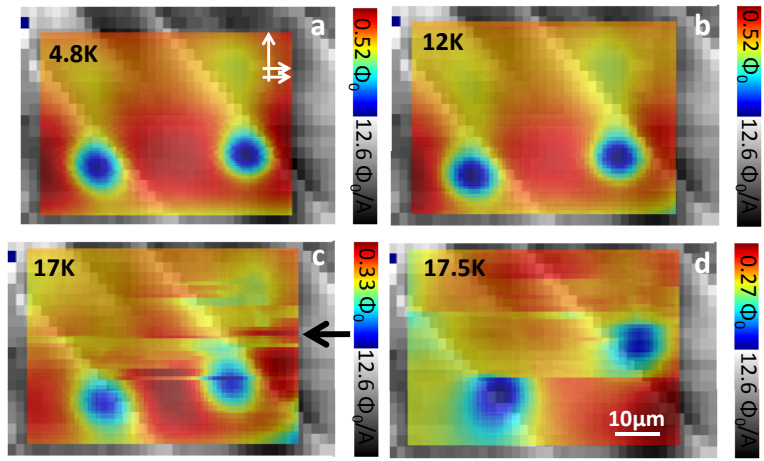

Figure 6